\pdfoutput=1
\documentclass[%
preprint,
 amsmath,amssymb,
 aps,
pra,
floatfix,
]{revtex4-2}
\usepackage{graphicx}
\usepackage{dcolumn}
\usepackage{bm}

\usepackage{multirow}
\usepackage{graphicx}

\begin{document}

\title{Microscopic origin of polarization-entangled Stokes--anti-Stokes photons in diamond}

\author{Tiago A. Freitas$^{1}$}
\author{Paula Machado$^{2,3}$}
\author{Lucas V. de Carvalho$^2$}
\author{Diego Sier$^{2}$}
\author{Raul Corrêa$^{2}$}
\author{Riichiro Saito$^{4}$}
\author{Marcelo F. Santos $^{5}$}
\author{Carlos H. Monken $^2$}
\author{Ado Jorio$^{1,2}$}

\affiliation{$^1$Programa de Pós-Graduação em Engenharia Elétrica, $^2$Departamento de Física, Universidade Federal de Minas Gerais, Belo Horizonte, MG 30123-970, Brazil}
\affiliation{$^3$Instituto D'Or de Pesquisa e Ensino, Rio de Janeiro, RJ 22281-100, Brazil}
\affiliation{$^4$Department of Physics, Tohoku University, Sendai, 980-8578, Japan}
\affiliation{$^5$Instituto de Física, UFRJ, Caixa Postal 68528, Rio de Janeiro, RJ 21941-972, Brazil}

\date{\today}

\begin{abstract}
Violation of the Clauser-Horne-Shimony-Holt inequality for the polarization of Stokes-anti-Stokes (SaS) photon pairs near a Raman resonance is demonstrated. The pairs are generated by shining a pulsed laser on a diamond sample, where two photons of the laser are converted into a pair of photons of different frequencies. The generated pairs are collected by standard Bell analyzers and shown to be entangled in polarization, with the degree of entanglement depending on the spectral region and on the orientation of the polarization of the incident light with respect to the crystallographic orientation of the sample. This result opens up the possibility to combine quantum optics and SaS Raman spectroscopy in order to improve materials science and quantum information.
\end{abstract}

\maketitle


\par
\begin{figure}[htp!]
	\centering
{\includegraphics[width=0.35\textwidth]{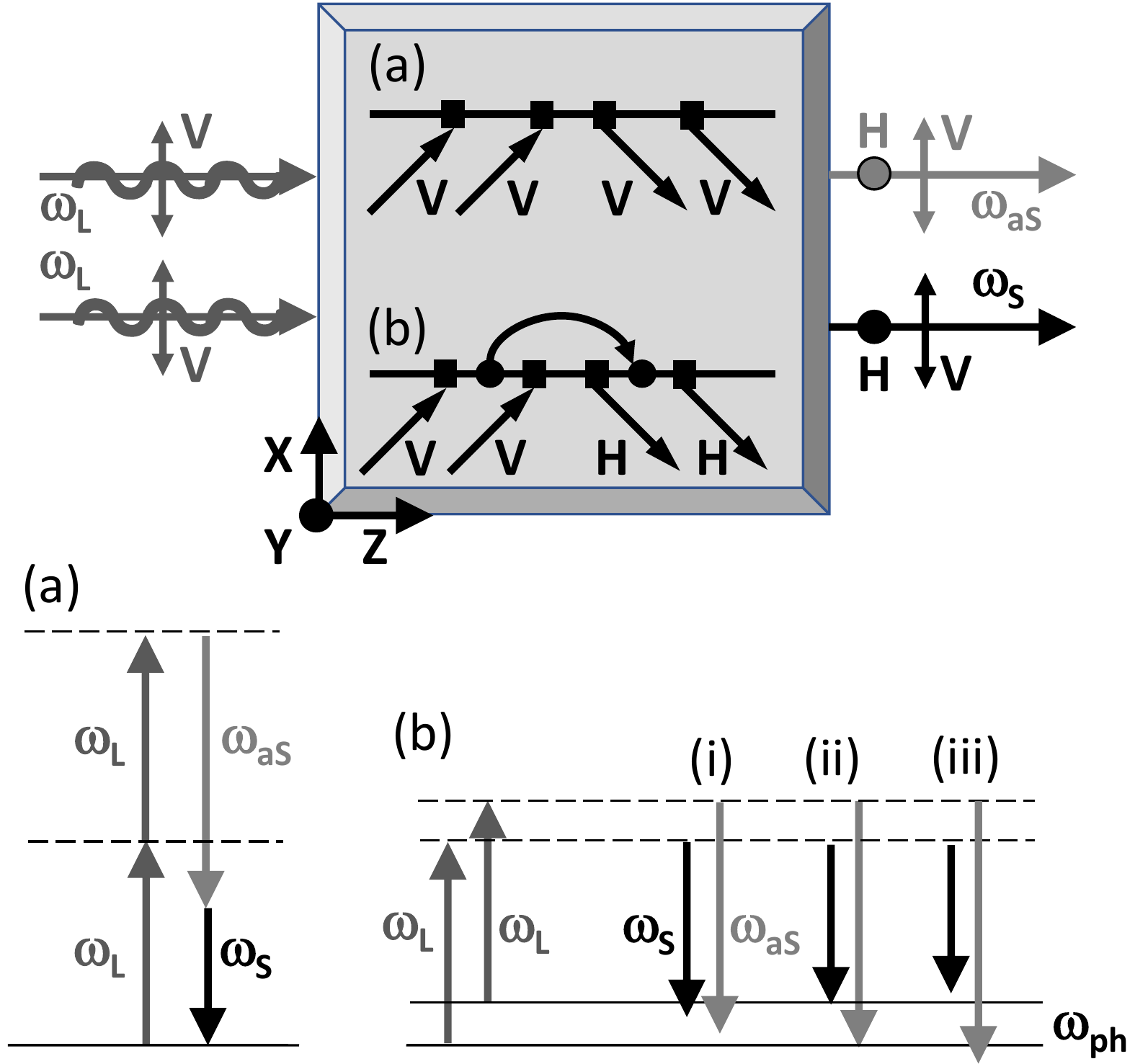}}

	\caption{Microscopic model evidencing the two paths where two photons from a laser ($\omega_L$) can generate an SaS pair ($\omega_S$ and $\omega_{aS}$): (a) purely electronic four-wave mixing; (b) phonon mediated process, which can be resonant (ii) or detuned out of resonance: (i) negative detuning; (iii) positive detuning. V and H are the orthogonal light polarizations. X,Y,Z are the diamond crystallographic axes.}
	\label{fig:model}
\end{figure}

 Raman scattering is the inelastic scattering of light by matter, where incident photons lose (Stokes) or gain (anti-Stokes) energy from the material medium.
Most commonly the energy exchanged between the photon and the medium is via a quantum of atomic vibration, i.e. a phonon. Of particular interest is the phenomenon where the same phonon generated in the Stokes process is absorbed in an anti-Stokes process, generating a pair of quantum correlated Stokes--anti-Stokes (SaS) photons \cite{klyshko1977correlation,Parra-Murillo2016,guimaraes2020stokes,kishore2021}. This process is a special type of degenerate four-wave mixing, where two photons of the same energy are changed into two photons of different energies inside the material medium, fulfilling energy conservation, $2\hbar\omega_{L} = \hbar\omega_{S} + \hbar\omega_{aS}$ (see Fig.\,\ref{fig:model}) \cite{bloembergennonlinear}. The four-wave mixing process can have a purely electronic microscopic origin [Fig.\,\ref{fig:model}(a)], or it can be mediated by a phonon [Fig.\,\ref{fig:model}(b)]. The phonon-mediated process is clearly distinguishable when resonance is obtained [Fig.\,\ref{fig:model}(b)(ii)], i.e. when $\omega_{S,aS} = \omega_{L}\pm \omega_{ph}$, where the $\pm$ stands for the anti-Stokes/Stokes scattering and $\hbar\omega_{ph}$ is the phonon energy \cite{levenson1972,levenson1974}. When tuned out of resonance [Fig.\,\ref{fig:model}(b)(i) and (iii)], the phononic SaS pairing can still be mediated by a virtual phonon, having, in this case, an intriguing parallel with the BCS theory for superconductivity \cite{Saraiva2017,Junior2019,zhang2018generation,de2019physical}.

\par Correlated photon pairs have attracted great interest in quantum optics and represent one of the most explored resources in experimental quantum optics and applications, with significant dissemination in several related research fields \cite{eisaman2011,senelart2017,caspani2017}. Discrete sources of non-classic correlated photon pairs resulting from a mixture of four waves have been performed on optical waveguides, photonic fibers, or nonlinear crystals \cite{chen2014,li2019,vergyris2020,llewellyn2020}.

\par In condensed matter physics, the uncorrelated Stokes to anti-Stokes intensity ratio gives the local temperature \cite{Parra-Murillo2016}. In high fields, however, the Stokes--anti-Stokes correlation generates rich phenomena in the light-matter interaction that has been largely explored \cite{kneipp1996population,brolo2004ratio,le2006vibrational,roelli2016molecular,schmidt2016quantum,zhang2020optomechanical}, including their uses to create quantum information transmission protocols \cite{Kuzmich2003a}, such as quantum optical memory \cite{England2013}, and to study one phonon Fock states \cite{velez2019}.

This phenomenon is general, observed in molecular systems like transparent liquids \cite{Saraiva2017,vento2023measurement}, including water \cite{Kasperczyk2016}, and in non-vibrational related Raman scattering, e.g. in rubidium \cite{Bashkansky2012}, cesium vapor \cite{Reim2010}, hydrogen molecules \cite{bustard2015}, among others. Diamond is, however, the preferred material to study such phenomena \cite{Lee2012,kasperczyk2015stokes,anderson2018,velez2019,Junior2019,Junior2020}, where the dependence on excitation laser power \cite{kasperczyk2015stokes,Junior2019}, Raman shift and scattering momenta \cite{Junior2019}, and lifetimes for both resonant (real phonon) and non-resonant (virtual phonon) processes \cite{Junior2020} have already been explored. While the non-classical nature of the SaS photons in diamond is well established, entanglement has been demonstrated only when two interfering pathways are engineered, such as two different samples \cite{Lee2011a},  time bin with two different pulses \cite{velez2020}, or two spin states within the same sample \cite{dolde2013}. What is missing is a report on the presence of entanglement in the fundamental four-wave mixing process in diamond.
Here, we demonstrate a pair of photons entangled through the superposition of the two distinct microscopic generation processes. In our experiment, the electronic degenerate four-wave mixing and vibrational SaS Raman scattering shown in Fig.\,\ref{fig:model}(a,b) collaborate to produce pairs of photons entangled in polarization, verified using the Clauser-Horne-Shimony-Holt (CHSH) inequality. Interestingly, we found that the condition of entanglement of the photons, in the case of diamond, depends on the spectral region and crystallographic orientation of the sample.


\begin{figure}[htp!]
	\centering
	{\includegraphics[width=0.45\textwidth]{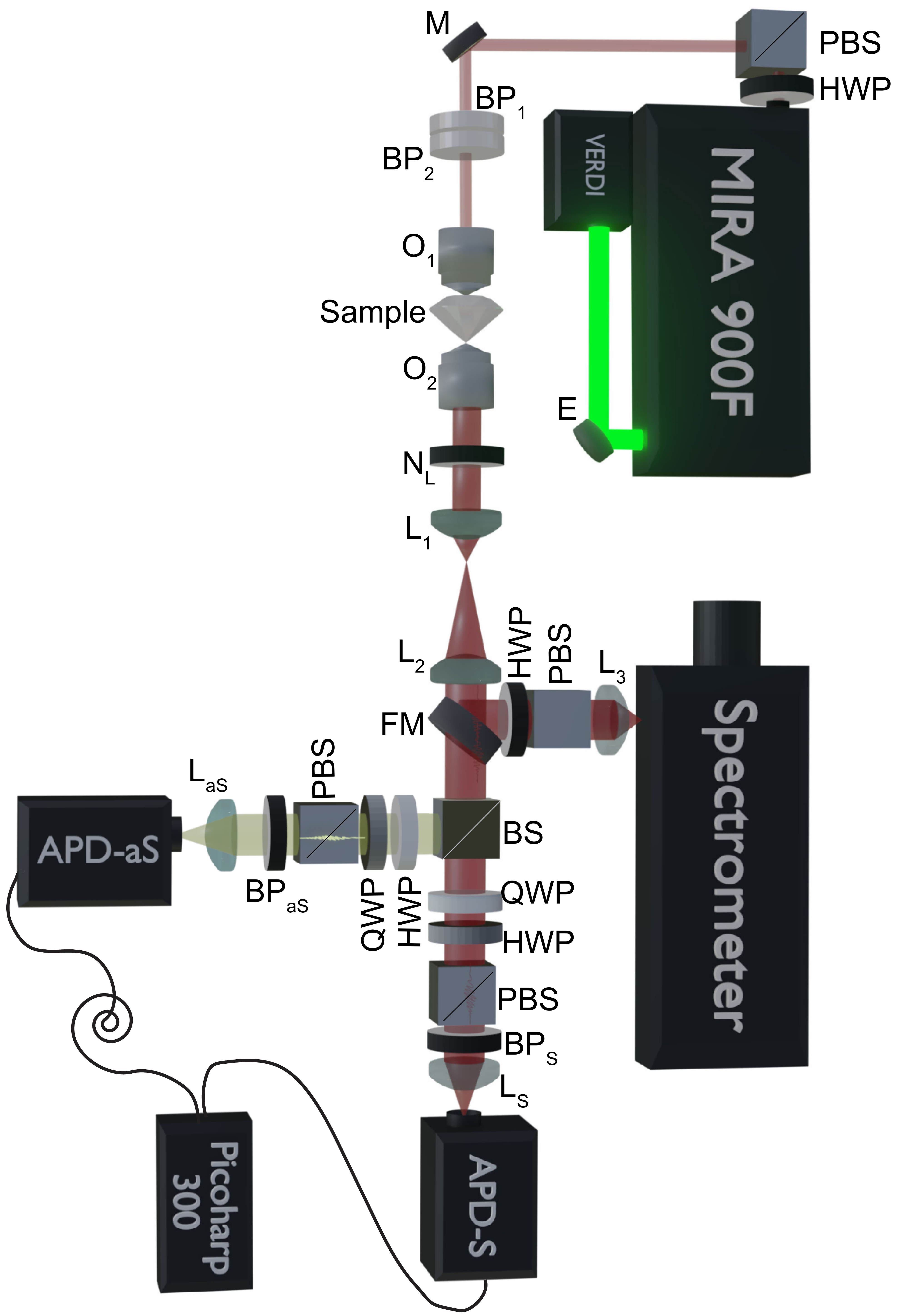}}
	\caption{Schematics of the experimental setup. (E) dielectric mirror for high-power, (HWP) half-wave plate, (PBS) polarized beamsplitter, (M) silver mirror, (BP$_{1,2}$) interference filter band-pass, (O$_1$) objective responsible for focusing the excitation laser (20$\times$, NA 0.5), (O$_2$) objective responsible for collecting the scattered signal (100$\times$, NA 0.9), (N$_L$ notch filter), (L$_{1,2,3,S,aS}$) plano-convex achromatic lenses, (FM) flip mirror, (BS) 50/50 beam-splitter, (QWP) quarter-wave plate, (BP$_{\!S,aS}$) bandpass interference filter to Stokes,anti-Stokes.}
	\label{fig:setuptomografia}
\end{figure}

Figure \ref{fig:setuptomografia} illustrates the experimental system utilized, which combines a Raman spectrometer with the time-correlated single-photon counting (TCSPC) technique. 
A 200\,fs pulsed MIRA 900F laser at 785\,nm and 76 MHz was used, powered by a Verdi G10 laser from Coherent. 
The laser with polarization horizontal ($H$) with respect to the laboratory reference frame departing from the MIRA 900F passes through a half-wave plate (HWP) and a polarized beamsplitter (PBS) to control power and convert polarization to vertical ($V$) before reaching the sample. With the flip mirror (FM in Fig.\,\ref{fig:setuptomografia}) in place, the sample can be spectrally characterized through conventional Raman spectroscopy. To investigate the correlation of single photons coming from the sample, the flip mirror (FM) is removed and the scattered light is sent to a 50/50 beamsplitter (BS). After the BS, a half-wave plate (HWP), a quarter-wave plate (QWP) and a polarized beamsplitter (PBS) are used to select the polarizations of the $S$ and $aS$ photons. After the PBS, a bandpass interference filter (BP$_{aS}$) centered on $\omega_{aS}$ and a BP$_{S}$ centered on $\omega_{S}$ are used for selecting the desired signals. The filtered signals are focused on avalanche photodiode detectors (APDs) Excelitas model (SPCM-AQRH-14) using plano-convex achromatic lenses (L$_{S}$ and L$_{aS}$).

The sample we use is a highly pure diamond grown by the CVD process (Type IIac Diacell design(100) oriented), with the laser propagating along the [001] direction of the diamond crystal. The sample is mounted on a rotation stage so that the angle $\theta$ between the diamond crystallographic axes and the incident laser polarization can be varied. $\theta = 0^{\circ}$ stands for the sample [100] direction along V and $\theta = 90^{\circ}$ stands for the sample [100] direction along $H$ (or the [010]-direction along $V$).

\begin{figure}[htp!]
	\centering
 {\includegraphics[width=0.49\textwidth]{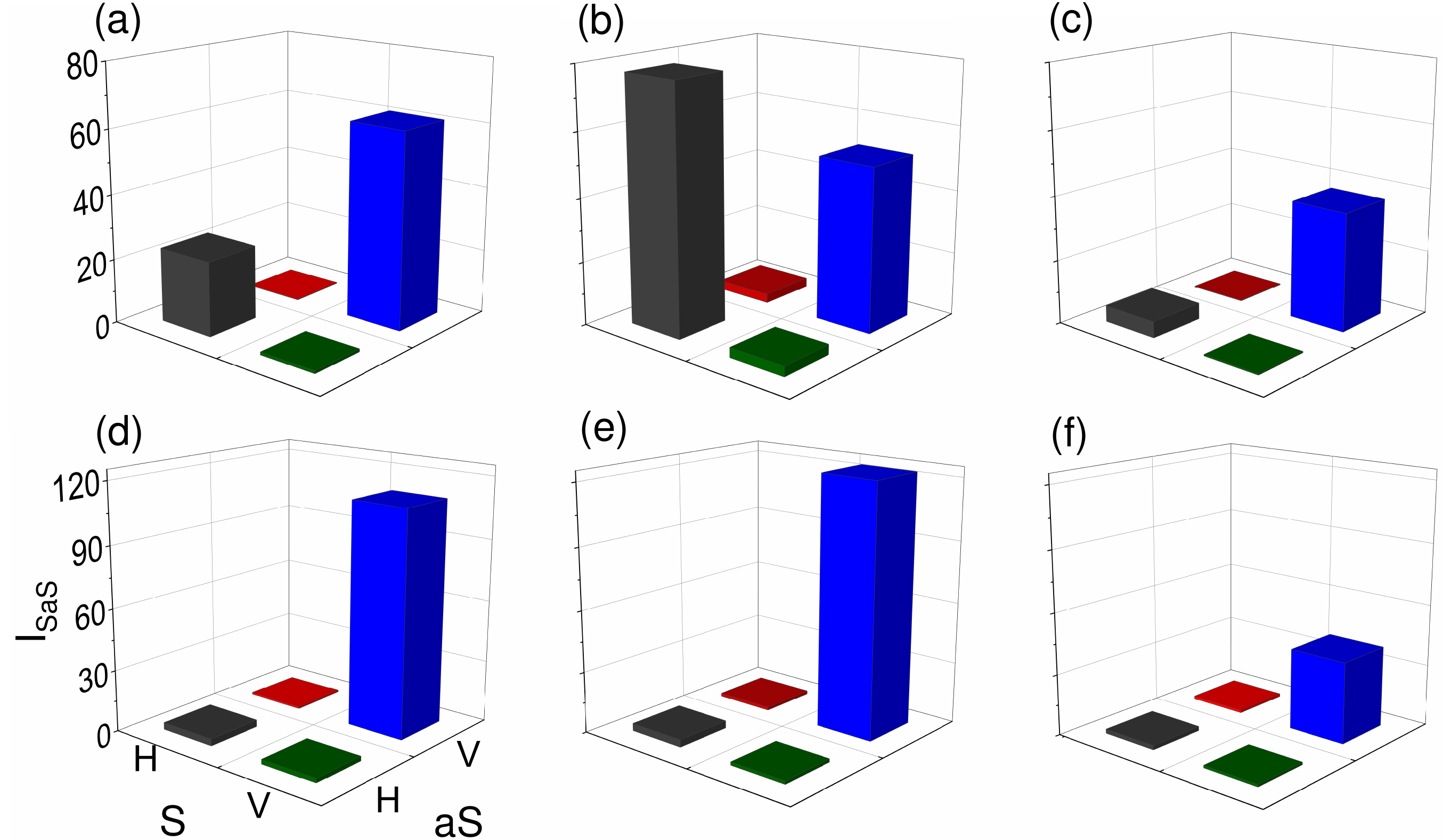}}
	\caption{Time-correlated SaS photon pair intensity ($I_{SaS}$) at three spectral regions: (a,d) $|\omega_{S,aS}| = 900$\,cm$^{-1}$ (S filter: 850/10; aS filter: 730/10); (b,e) $|\omega_{S,aS}| = 1332$\,cm$^{-1}$ (S: 875/25; aS: 711/25); (c,f) $|\omega_{S,aS}| = 1900$\,cm$^{-1}$ (S: 920/10; aS: 685/40). Laser excitation at 785\,nm. (a,b,c) $\theta= 0^{\circ}$, (d,e,f) $\theta= 45^{\circ}$. $I_{SaS}$ were corrected to account for the spectral efficiency of the experimental apparatus and different excitation powers. The scales of $I_{SaS}$ are in counts/cm$^{-1}\cdot$mW$^{2}\cdot$s.}
	\label{fig:diagonal}

\end{figure}

Figure \ref{fig:diagonal} represents the results for time-correlated SaS Raman scattering ($I_{SaS}$) at three spectral regions: (a,d) detuned from resonance with $|\omega_{S,aS}|<\omega_{ph}$; (b,e) in resonance with the phonon, $|\omega_{S,aS}|=\omega_{ph}$; and (c,f) detuned from resonance with $|\omega_{S,aS}|>\omega_{\rm ph}$. The top row (a,b,c) stands for $\theta= 0^{\circ}$ and the bottom row (d,e,f) for $\theta= 45^{\circ}$, where for $\theta = 0^{\circ}$, $V = X$ in Fig.\,\ref{fig:model}, while for $\theta = 45^{\circ}$, $V = X+Y$, which are the two inequivalent high symmetry directions in the $XY$ plane in the cubic diamond crystal. Light is propagating along $Z$. The observed correlation functions $g^2(\tau = 0)$ always exceed the classical limit 2, indicating quantum correlated signals.

As shown in Fig.\,\ref{fig:diagonal}, in all scattering configurations, the SaS photon pair is always observed with the same polarization, i.e. either in $V_SV_{aS}$ or $H_SH_{aS}$, with practically no signal being observed in $V_SH_{aS}$ or $H_SV_{aS}$. For $\theta= 0^{\circ}$ (top row), both $V_SV_{aS}$ and $H_SH_{aS}$ are seen, while for $\theta= 45^{\circ}$ (bottom row) only $V_SV_{aS}$ signal can be seen.
When the polarization of the incident laser is oriented along the diamond crystallographic axis $\theta=0^{\circ}$, Fig.\ref{fig:diagonal}(a,b,c), the statistics suggests that the SaS state is of the type
\begin{equation}
\label{eq:SaSstate}
|S,aS\rangle =c_{1} |V_S,V_{aS}\rangle + c_{2} |H_S,H_{aS}\rangle.
\end{equation}
Notice that the amount of $H_SH_{aS}$ signal varies much more than the amount of $V_{S}V_{aS}$ signal when changing the spectral region, $H_{S}H_{aS}$ being significantly larger in resonance. Comparing the counts in the top and bottom rows, it is clear that for $\theta= 45^{\circ}$, the $H_SH_{aS}$ signal observed in $\theta= 0^{\circ}$ is converted to $V_SV_{aS}$.

For the scattering configuration of Fig.\,\ref{fig:diagonal}(a), the values of $c_1$ and $c_2$ in Eq.\,\ref{eq:SaSstate} are $\sqrt{0.72}$ and $\sqrt{0.28}$, i.e. $|S,aS\rangle$ is very close to the $|\phi^+\rangle$ Bell state (more precisely: $\sqrt{0.95}|\phi^+\rangle + \sqrt{0.05}|\phi^-\rangle$) [where $|\phi^{\pm}\rangle = (|VV\rangle \pm |HH\rangle)/\sqrt{2}$]. Therefore, in order to validate the entanglement of the SaS state, we performed polarization correlation measurements that allowed us to test the CHSH inequality that is maximally violated by $|\phi^+\rangle$. The predicted value of the CHSH parameter $|S|$ in this case is $|S|=2.68$.
After measurements in- and out of-resonance, for $\theta = 0^{\circ}$ and $45^{\circ}$, and after data treatment, the obtained CHSH parameters are displayed in Table\,\ref{tab:bellcalc}.

\begin{table}[hbt!]
\caption{\label{tab:bellcalc}%
Bell-CHSH parameter $|S|$ estimation for the SaS-state in- and out-resonance, for the diamond orientations $\theta = 0^{\circ}$ and $45^{\circ}$.}
\begin{ruledtabular}
\begin{tabular}{cccc}
$\theta$ & 900\,cm$^{-1}$& 1332\,cm$^{-1}$ & 1900\,cm$^{-1}$\\
\colrule
$0^\circ$ & 2.61$\pm$0.03 & 2.05$\pm$0.16 & 0.24$\pm$0.16 \\
$45^\circ$   &1.82$\pm$0.03  & 1.62$\pm$0.17 & 1.82$\pm$0.07 \\
\end{tabular}
\end{ruledtabular}
\end{table}

The violation of the CHSH inequality for the virtual-phonon scattering with the crystal oriented along the incident laser polarization ($\theta = 0^{\circ}$) is clear for the $900$ cm$^{-1}$ pairs, in good agreement with the predicted value of 2.68. For the real scattering, on the other hand, the violation is marginal, likely due to the presence of a much larger background of uncorrelated photons. The downgrade of the correlation at the Raman resonance results from a statistical mixture of the SaS entangled pairs with unentangled three-photon states. Three photon states are mainly due to the sum of a standard single photon Stokes scattering and a SaS pair. The detectors do not distinguish between two and three-photon events and the measurement apparatus ends up bunching them in the same correlation statistics. This behavior was already analyzed in previous papers as the physical origin of the severe lowering of the value of the crossed zero-time correlation function $g^2_{SaS}(0)$ at the Raman resonance~\cite{kasperczyk2015stokes,de2019physical}. We have also confirmed it by performing the tomography of the post-selected two-photon states, obtaining a purity ($\textrm{Tr}[\rho^2]$) of around 0.65, which indicates a very mixed two-qubit state for the polarizations of the SaS pair. For the $\theta=45^{\circ}$ orientation, there is no CHSH violation because essentially all the photons are generated at the same polarization. We also do not see any violation for the $1900$ cm$^{-1}$ pairs because the SaS counts are too low and the entangled pairs are of the same order of the accidental counts \cite{Junior2019,de2019physical}. At this spectral range the overall polarization state is always essentially $|VV\rangle$, no matter the crystalographic orientation.


The obtained results for the entangled SaS pairs at 900 cm$^{-1}$ can be explained considering that the SaS polarization state $\left|S,aS\right>$ is given by 
\begin{equation}
\left|S,aS\right> = \left[\left(\chi^e\otimes\chi^e\right)\oplus\left(\alpha^R\otimes\alpha^R\right)\right]\left|L\right>\otimes\left|L\right>\, ,
\label{Eq:SaS1}
\end{equation}
where $\chi^e$ is the electronic susceptibility, $\alpha^R$ is the Raman polarizability tensor and $\left|L\right>$ is the laser polarization state. Since diamond is a cubic system ($O_h$ point symmetry group), $\chi^e$ is a diagonal tensor with one independent tensor element, and $\alpha^R$ belongs to the triple degenerate $T_{2g}$ irreducible representation. For both incident and scattered light propagating along [001], only one Raman tensor applies
\begin{equation}
\alpha^R =
\left(
\begin{tabular}{lll}
0 & 1 & 0 \\
1 & 0 & 0 \\
0 & 0 & 0
 \end{tabular}
\right).
\label{Eq:T2g}
\end{equation}
For the incident laser polarized vertically ($V$) with respect to the laboratory reference frame, i.e. $\left|L\right>\otimes\left|L\right> = \left|V_L\right>\otimes\left|V_L\right>$, the scattered SaS state will be either $\left|S,aS\right> = \left|V_S,V_{aS}\right>$ or $\left|S,aS\right> = \left|H_S,H_{aS}\right>$ depending on the electronic versus vibronic scattering path, and also depending on the crystallographic orientation of the diamond sample. If the sample is oriented with the crystallographic axes rotated by $\pi/4$ with respect to the laboratory reference axes, then both the Raman and electronic paths generate $\left|V_SV_{aS}\right>$. However, if the diamond sample is oriented with the crystallographic axes aligned with the laboratory reference axes, then the Raman path generates $\left|H_S,H_{aS}\right>$ while the electronic path generates $\left|V_S,V_{aS}\right>$. In the later case, the generated state is $\left|S,aS\right> = c_1\left|V_S,V_{aS}\right> + c_2\left|H_S,H_{aS}\right> = c_1\left|\chi^e_S,\chi^e_{aS}\right> + c_2\left|\alpha^R_S,\alpha^R_{aS}\right>$, where $\chi^e$ stands for the electronic path and $\alpha^R$ stands for the Raman (vibrational) path. This is an entangled state where the process can be either electronic or vibronic, as pictured in Fig.\,\ref{fig:model}. Notice, in Fig.\,\ref{fig:diagonal}(a,b,c), that while the electronic four-wave mixing ($VV$) decays monotonically while scanning the selected spectral region through resonance, the Raman part ($HH$) strongly enhances when in resonance ($|\omega_{S,aS}| = 1332$\,cm$^{-1}$), and significantly decays for positive detuning, responsible for the previously observed ($\pm$)-detuning asymmetry \cite{Junior2019,de2019physical}.


To conclude, we have demonstrated the production of polarization-entangled photon pairs in a diamond sample through a four-wave mixing process. The entanglement is obtained only when the crystal orientation favors two distinct paths for generating the four-wave mixing. For light reaching the sample in the [100] direction, entanglement is obtained when the incident laser is polarized along one of the crystallographic axes, since electronic four-wave mixing keeps the polarization while vibronic four-wave mixing (SaS scattering) changes the polarization. As a result, the superposition of the two possible processes generates pairs of photons in the entangled polarization state $\left|S,aS\right> = c_1\left|\chi^e_S,\chi^e_{aS}\right> + c_2\left|\alpha^R_S,\alpha^R_{aS}\right>$.

Intriguingly, to excite an electron in diamond, it takes 5.4\,eV, while to excite an optical phonon, it takes 0.14\,eV. Consequently, the adiabatic approximation is utilized to handle electrons and phonons separately. However, here we show that it is possible to entangle the electronic and nuclear motion in diamond when we make light travel in the crystal in well-defined geometries, so that the light-matter interaction happens in a way that one cannot, in advance, tell which process will take place.

From another perspective, for light reaching the diamond sample in the [110] direction, it would be possible to generate entanglement from two different degenerate phonon paths, since the $T_{2g}$ phonon in diamond is triple-degenerate and the specific partner can be selected by a corresponding specific scattered light polarization.

These results open up the possibility to combine quantum information and SaS Raman spectroscopy in order to improve techniques in both areas, as well as to create new ones. As an immediate example, one can think of using entangled SaS light as a tunable source of different color photon pairs for quantum communication and quantum probing. A more promising outcome, however, may lie in a deeper integration of quantum information protocols into Raman spectroscopy, allowing for a potentially large expansion of an already powerful method to probe materials and biological systems.

\section*{Acknowledgments}
The authors acknowledge
finantial support by Fapemig (TEC-RED-00282-16, APQ-01860-22), CNPq (302775/2018-8, 442521/2019-7, 313158/2022-3), FAPERJ (E-26/202.576/2019, E-26/200.307/2023), CAPES-PRINT/UFMG and JSPS Kakenhi (Nos. JPI8H01810, JP22H00283).


\begin{thebibliography}{51}%
	\makeatletter
	\providecommand \@ifxundefined [1]{%
	 \@ifx{#1\undefined}
	}%
	\providecommand \@ifnum [1]{%
	 \ifnum #1\expandafter \@firstoftwo
	 \else \expandafter \@secondoftwo
	 \fi
	}%
	\providecommand \@ifx [1]{%
	 \ifx #1\expandafter \@firstoftwo
	 \else \expandafter \@secondoftwo
	 \fi
	}%
	\providecommand \natexlab [1]{#1}%
	\providecommand \enquote  [1]{``#1''}%
	\providecommand \bibnamefont  [1]{#1}%
	\providecommand \bibfnamefont [1]{#1}%
	\providecommand \citenamefont [1]{#1}%
	\providecommand \href@noop [0]{\@secondoftwo}%
	\providecommand \href [0]{\begingroup \@sanitize@url \@href}%
	\providecommand \@href[1]{\@@startlink{#1}\@@href}%
	\providecommand \@@href[1]{\endgroup#1\@@endlink}%
	\providecommand \@sanitize@url [0]{\catcode `\\12\catcode `\$12\catcode `\&12\catcode `\#12\catcode `\^12\catcode `\_12\catcode `\%12\relax}%
	\providecommand \@@startlink[1]{}%
	\providecommand \@@endlink[0]{}%
	\providecommand \url  [0]{\begingroup\@sanitize@url \@url }%
	\providecommand \@url [1]{\endgroup\@href {#1}{\urlprefix }}%
	\providecommand \urlprefix  [0]{URL }%
	\providecommand \Eprint [0]{\href }%
	\providecommand \doibase [0]{https://doi.org/}%
	\providecommand \selectlanguage [0]{\@gobble}%
	\providecommand \bibinfo  [0]{\@secondoftwo}%
	\providecommand \bibfield  [0]{\@secondoftwo}%
	\providecommand \translation [1]{[#1]}%
	\providecommand \BibitemOpen [0]{}%
	\providecommand \bibitemStop [0]{}%
	\providecommand \bibitemNoStop [0]{.\EOS\space}%
	\providecommand \EOS [0]{\spacefactor3000\relax}%
	\providecommand \BibitemShut  [1]{\csname bibitem#1\endcsname}%
	\let\auto@bib@innerbib\@empty
	\bibitem [{\citenamefont {Klyshko}(1977)}]{klyshko1977correlation}%
		\BibitemOpen
		\bibfield  {author} {\bibinfo {author} {\bibfnamefont {D.~N.}\ \bibnamefont {Klyshko}},\ }\bibfield  {title} {\bibinfo {title} {{Correlation between the Stokes and anti-Stokes components in inelastic scattering of light}},\ }\href {https://doi.org/10.1070/qe1977v007n06abeh012890} {\bibfield  {journal} {\bibinfo  {journal} {Soviet Journal of Quantum Electronics}\ }\textbf {\bibinfo {volume} {7}},\ \bibinfo {pages} {755} (\bibinfo {year} {1977})}\BibitemShut {NoStop}%
	\bibitem [{\citenamefont {Parra-Murillo}\ \emph {et~al.}(2016)\citenamefont {Parra-Murillo}, \citenamefont {Santos}, \citenamefont {Monken},\ and\ \citenamefont {Jorio}}]{Parra-Murillo2016}%
		\BibitemOpen
		\bibfield  {author} {\bibinfo {author} {\bibfnamefont {C.~A.}\ \bibnamefont {Parra-Murillo}}, \bibinfo {author} {\bibfnamefont {M.~F.}\ \bibnamefont {Santos}}, \bibinfo {author} {\bibfnamefont {C.~H.}\ \bibnamefont {Monken}},\ and\ \bibinfo {author} {\bibfnamefont {A.}~\bibnamefont {Jorio}},\ }\bibfield  {title} {\bibinfo {title} {Stokes--anti-Stokes correlation in the inelastic scattering of light by matter and generalization of the Bose-Einstein population function},\ }\href@noop {} {\bibfield  {journal} {\bibinfo  {journal} {Physical Review B}\ }\textbf {\bibinfo {volume} {93}},\ \bibinfo {pages} {125141} (\bibinfo {year} {2016})}\BibitemShut {NoStop}%
	\bibitem [{\citenamefont {Guimar{\~a}es}\ \emph {et~al.}(2020)\citenamefont {Guimar{\~a}es}, \citenamefont {Santos}, \citenamefont {Jorio},\ and\ \citenamefont {Monken}}]{guimaraes2020stokes}%
		\BibitemOpen
		\bibfield  {author} {\bibinfo {author} {\bibfnamefont {A.~V.~A.}\ \bibnamefont {Guimar{\~a}es}}, \bibinfo {author} {\bibfnamefont {M.~F.}\ \bibnamefont {Santos}}, \bibinfo {author} {\bibfnamefont {A.}~\bibnamefont {Jorio}},\ and\ \bibinfo {author} {\bibfnamefont {C.~H.}\ \bibnamefont {Monken}},\ }\bibfield  {title} {\bibinfo {title} {Stokes--anti-Stokes light-scattering process: a photon-wave-function approach},\ }\href@noop {} {\bibfield  {journal} {\bibinfo  {journal} {Physical Review A}\ }\textbf {\bibinfo {volume} {102}},\ \bibinfo {pages} {033719} (\bibinfo {year} {2020})}\BibitemShut {NoStop}%
	\bibitem [{\citenamefont {Thapliyal}\ and\ \citenamefont {Pe{\v{r}}ina~Jr}(2021)}]{kishore2021}%
		\BibitemOpen
		\bibfield  {author} {\bibinfo {author} {\bibfnamefont {K.}~\bibnamefont {Thapliyal}}\ and\ \bibinfo {author} {\bibfnamefont {J.}~\bibnamefont {Pe{\v{r}}ina~Jr}},\ }\bibfield  {title} {\bibinfo {title} {Ideal pairing of the Stokes and anti-Stokes photons in the Raman process},\ }\href@noop {} {\bibfield  {journal} {\bibinfo  {journal} {Physical Review A}\ }\textbf {\bibinfo {volume} {103}},\ \bibinfo {pages} {033708} (\bibinfo {year} {2021})}\BibitemShut {NoStop}%
	\bibitem [{\citenamefont {Bloembergen}(1974)}]{bloembergennonlinear}%
		\BibitemOpen
		\bibfield  {author} {\bibinfo {author} {\bibfnamefont {N.}~\bibnamefont {Bloembergen}},\ }\href@noop {} {\emph {\bibinfo {title} {Nonlinear Optics}}}\ (\bibinfo  {publisher} {W. A. Benjamin, New York},\ \bibinfo {year} {1974})\BibitemShut {NoStop}%
	\bibitem [{\citenamefont {Levenson}\ \emph {et~al.}(2021)\citenamefont {Levenson}, \citenamefont {Flytzanis},\ and\ \citenamefont {Bloembergen}}]{levenson1972}%
		\BibitemOpen
		\bibfield  {author} {\bibinfo {author} {\bibfnamefont {M.~D.}\ \bibnamefont {Levenson}}, \bibinfo {author} {\bibfnamefont {C.}~\bibnamefont {Flytzanis}},\ and\ \bibinfo {author} {\bibfnamefont {N.}~\bibnamefont {Bloembergen}},\ }\bibfield  {title} {\bibinfo {title} {{Interference of Resonant and Nonresonant Three-Wave Mixing in Diamond}},\ }\href@noop {} {\bibfield  {journal} {\bibinfo  {journal} {Physical Review B}\ }\textbf {\bibinfo {volume} {6}},\ \bibinfo {pages} {103962} (\bibinfo {year} {1972})}\BibitemShut {NoStop}%
	\bibitem [{\citenamefont {Levenson}\ and\ \citenamefont {Bloembergen}(1974)}]{levenson1974}%
		\BibitemOpen
		\bibfield  {author} {\bibinfo {author} {\bibfnamefont {M.~D.}\ \bibnamefont {Levenson}}\ and\ \bibinfo {author} {\bibfnamefont {N.}~\bibnamefont {Bloembergen}},\ }\bibfield  {title} {\bibinfo {title} {Dispersion of the nonlinear optical susceptibility tensor in centrosymmetric media},\ }\href {https://doi.org/10.1103/PhysRevB.10.4447} {\bibfield  {journal} {\bibinfo  {journal} {Phys. Rev. B}\ }\textbf {\bibinfo {volume} {10}},\ \bibinfo {pages} {4447} (\bibinfo {year} {1974})}\BibitemShut {NoStop}%
	\bibitem [{\citenamefont {Saraiva}\ \emph {et~al.}(2017)\citenamefont {Saraiva}, \citenamefont {J{\'u}nior}, \citenamefont {de~Melo~e Souza}, \citenamefont {Pena}, \citenamefont {Monken}, \citenamefont {Santos}, \citenamefont {Koiller},\ and\ \citenamefont {Jorio}}]{Saraiva2017}%
		\BibitemOpen
		\bibfield  {author} {\bibinfo {author} {\bibfnamefont {A.}~\bibnamefont {Saraiva}}, \bibinfo {author} {\bibfnamefont {F.~S.}\ \bibnamefont {de~Aguiar~J{\'u}nior}}, \bibinfo {author} {\bibfnamefont {R.}~\bibnamefont {de~Melo~e Souza}}, \bibinfo {author} {\bibfnamefont {A.~P.}\ \bibnamefont {Pena}}, \bibinfo {author} {\bibfnamefont {C.~H.}\ \bibnamefont {Monken}}, \bibinfo {author} {\bibfnamefont {M.~F.}\ \bibnamefont {Santos}}, \bibinfo {author} {\bibfnamefont {B.}~\bibnamefont {Koiller}},\ and\ \bibinfo {author} {\bibfnamefont {A.}~\bibnamefont {Jorio}},\ }\bibfield  {title} {\bibinfo {title} {Photonic counterparts of Cooper pairs},\ }\href@noop {} {\bibfield  {journal} {\bibinfo  {journal} {Physical Review Letters}\ }\textbf {\bibinfo {volume} {119}},\ \bibinfo {pages} {193603} (\bibinfo {year} {2017})}\BibitemShut {NoStop}%
	\bibitem [{\citenamefont {J{\'u}nior}\ \emph {et~al.}(2019)\citenamefont {J{\'u}nior}, \citenamefont {Saraiva}, \citenamefont {Santos}, \citenamefont {Koiller}, \citenamefont {Souza}, \citenamefont {Pena}, \citenamefont {Silva}, \citenamefont {Monken},\ and\ \citenamefont {Jorio}}]{Junior2019}%
		\BibitemOpen
		\bibfield  {author} {\bibinfo {author} {\bibfnamefont {F.~S.}\ \bibnamefont {de~Aguiar~J{\'u}nior}}, \bibinfo {author} {\bibfnamefont {A.}~\bibnamefont {Saraiva}}, \bibinfo {author} {\bibfnamefont {M.~F.}\ \bibnamefont {Santos}}, \bibinfo {author} {\bibfnamefont {B.}~\bibnamefont {Koiller}}, \bibinfo {author} {\bibfnamefont {R.~d.~M.}\ \bibnamefont {Souza}}, \bibinfo {author} {\bibfnamefont {A.~P.}\ \bibnamefont {Pena}}, \bibinfo {author} {\bibfnamefont {R.~A.}\ \bibnamefont {Silva}}, \bibinfo {author} {\bibfnamefont {C.~H.}\ \bibnamefont {Monken}},\ and\ \bibinfo {author} {\bibfnamefont {A.}~\bibnamefont {Jorio}},\ }\bibfield  {title} {\bibinfo {title} {Stokes--anti-Stokes correlated photon properties akin to photonic Cooper pairs},\ }\href@noop {} {\bibfield  {journal} {\bibinfo  {journal} {Physical Review B}\ }\textbf {\bibinfo {volume} {99}},\ \bibinfo {pages} {100503(R)} (\bibinfo {year} {2019})}\BibitemShut {NoStop}%
	\bibitem [{\citenamefont {Zhang}\ \emph {et~al.}(2018)\citenamefont {Zhang}, \citenamefont {Zhang},\ and\ \citenamefont {Zhu}}]{zhang2018generation}%
		\BibitemOpen
		\bibfield  {author} {\bibinfo {author} {\bibfnamefont {Y.}~\bibnamefont {Zhang}}, \bibinfo {author} {\bibfnamefont {L.}~\bibnamefont {Zhang}},\ and\ \bibinfo {author} {\bibfnamefont {Y.-Y.}\ \bibnamefont {Zhu}},\ }\bibfield  {title} {\bibinfo {title} {Generation of photonic Cooper pairs in nanoscale optomechanical waveguides},\ }\href@noop {} {\bibfield  {journal} {\bibinfo  {journal} {Physical Review A}\ }\textbf {\bibinfo {volume} {98}},\ \bibinfo {pages} {013824} (\bibinfo {year} {2018})}\BibitemShut {NoStop}%
	\bibitem [{\citenamefont {de~Aguiar~J{\'u}nior}\ \emph {et~al.}(2019)\citenamefont {de~Aguiar~J{\'u}nior}, \citenamefont {Monken}, \citenamefont {Santos}, \citenamefont {de~Melo~e Souza}, \citenamefont {Saraiva}, \citenamefont {Koiller},\ and\ \citenamefont {Jorio}}]{de2019physical}%
		\BibitemOpen
		\bibfield  {author} {\bibinfo {author} {\bibfnamefont {F.~S.}\ \bibnamefont {de~Aguiar~J{\'u}nior}}, \bibinfo {author} {\bibfnamefont {C.~H.}\ \bibnamefont {Monken}}, \bibinfo {author} {\bibfnamefont {M.~F.}\ \bibnamefont {Santos}}, \bibinfo {author} {\bibfnamefont {R.}~\bibnamefont {de~Melo~e Souza}}, \bibinfo {author} {\bibfnamefont {A.}~\bibnamefont {Saraiva}}, \bibinfo {author} {\bibfnamefont {B.}~\bibnamefont {Koiller}},\ and\ \bibinfo {author} {\bibfnamefont {A.}~\bibnamefont {Jorio}},\ }\bibfield  {title} {\bibinfo {title} {Physical properties of photonic Cooper pairs generated via correlated Stokes--anti-Stokes Raman scattering},\ }\href@noop {} {\bibfield  {journal} {\bibinfo  {journal} {Physica Status Solidi (b)}\ }\textbf {\bibinfo {volume} {256}},\ \bibinfo {pages} {1900218} (\bibinfo {year} {2019})}\BibitemShut {NoStop}%
	\bibitem [{\citenamefont {Eisaman}\ \emph {et~al.}(2011)\citenamefont {Eisaman}, \citenamefont {Fan}, \citenamefont {Migdall},\ and\ \citenamefont {Polyakov}}]{eisaman2011}%
		\BibitemOpen
		\bibfield  {author} {\bibinfo {author} {\bibfnamefont {M.~D.}\ \bibnamefont {Eisaman}}, \bibinfo {author} {\bibfnamefont {J.}~\bibnamefont {Fan}}, \bibinfo {author} {\bibfnamefont {A.}~\bibnamefont {Migdall}},\ and\ \bibinfo {author} {\bibfnamefont {S.~V.}\ \bibnamefont {Polyakov}},\ }\bibfield  {title} {\bibinfo {title} {{Invited Review Article: Single-photon sources and detectors}},\ }\bibfield  {journal} {\bibinfo  {journal} {Review of Scientific Instruments}\ }\textbf {\bibinfo {volume} {82}},\ \href {https://doi.org/10.1063/1.3610677} {10.1063/1.3610677} (\bibinfo {year} {2011})\BibitemShut {NoStop}%
	\bibitem [{\citenamefont {Senellart}\ \emph {et~al.}(2017)\citenamefont {Senellart}, \citenamefont {Solomon},\ and\ \citenamefont {White}}]{senelart2017}%
		\BibitemOpen
		\bibfield  {author} {\bibinfo {author} {\bibfnamefont {P.}~\bibnamefont {Senellart}}, \bibinfo {author} {\bibfnamefont {G.}~\bibnamefont {Solomon}},\ and\ \bibinfo {author} {\bibfnamefont {A.}~\bibnamefont {White}},\ }\bibfield  {title} {\bibinfo {title} {{High-performance semiconductor quantum-dot single-photon sources}},\ }\href {https://doi.org/10.1038/nnano.2017.218} {\bibfield  {journal} {\bibinfo  {journal} {Nature Nanotechnology}\ }\textbf {\bibinfo {volume} {12}},\ \bibinfo {pages} {1026} (\bibinfo {year} {2017})}\BibitemShut {NoStop}%
	\bibitem [{\citenamefont {Caspani}\ \emph {et~al.}(2017)\citenamefont {Caspani}, \citenamefont {Xiong}, \citenamefont {Eggleton}, \citenamefont {Bajoni}, \citenamefont {Liscidini}, \citenamefont {Galli}, \citenamefont {Morandotti},\ and\ \citenamefont {Moss}}]{caspani2017}%
		\BibitemOpen
		\bibfield  {author} {\bibinfo {author} {\bibfnamefont {L.}~\bibnamefont {Caspani}}, \bibinfo {author} {\bibfnamefont {C.}~\bibnamefont {Xiong}}, \bibinfo {author} {\bibfnamefont {B.~J.}\ \bibnamefont {Eggleton}}, \bibinfo {author} {\bibfnamefont {D.}~\bibnamefont {Bajoni}}, \bibinfo {author} {\bibfnamefont {M.}~\bibnamefont {Liscidini}}, \bibinfo {author} {\bibfnamefont {M.}~\bibnamefont {Galli}}, \bibinfo {author} {\bibfnamefont {R.}~\bibnamefont {Morandotti}},\ and\ \bibinfo {author} {\bibfnamefont {D.~J.}\ \bibnamefont {Moss}},\ }\bibfield  {title} {\bibinfo {title} {{Integrated sources of photon quantum states based on nonlinear optics}},\ }\href {https://doi.org/10.1038/lsa.2017.100} {\bibfield  {journal} {\bibinfo  {journal} {Light: Science and Applications}\ }\textbf {\bibinfo {volume} {6}},\ \bibinfo {pages} {e17100} (\bibinfo {year} {2017})}\BibitemShut {NoStop}%
	\bibitem [{\citenamefont {Chen}\ \emph {et~al.}(2014)\citenamefont {Chen}, \citenamefont {Lei},\ and\ \citenamefont {Romero}}]{chen2014}%
		\BibitemOpen
		\bibfield  {author} {\bibinfo {author} {\bibfnamefont {L.}~\bibnamefont {Chen}}, \bibinfo {author} {\bibfnamefont {J.}~\bibnamefont {Lei}},\ and\ \bibinfo {author} {\bibfnamefont {J.}~\bibnamefont {Romero}},\ }\bibfield  {title} {\bibinfo {title} {{Quantum digital spiral imaging}},\ }\bibfield  {journal} {\bibinfo  {journal} {Light: Science and Applications}\ }\textbf {\bibinfo {volume} {3}},\ \href {https://doi.org/10.1038/lsa.2014.34} {10.1038/lsa.2014.34} (\bibinfo {year} {2014})\BibitemShut {NoStop}%
	\bibitem [{\citenamefont {Li}\ \emph {et~al.}(2019)\citenamefont {Li}, \citenamefont {Huang}, \citenamefont {Xiang}, \citenamefont {Nie}, \citenamefont {Sang}, \citenamefont {Yuan},\ and\ \citenamefont {Chen}}]{li2019}%
		\BibitemOpen
		\bibfield  {author} {\bibinfo {author} {\bibfnamefont {Y.}~\bibnamefont {Li}}, \bibinfo {author} {\bibfnamefont {Y.}~\bibnamefont {Huang}}, \bibinfo {author} {\bibfnamefont {T.}~\bibnamefont {Xiang}}, \bibinfo {author} {\bibfnamefont {Y.}~\bibnamefont {Nie}}, \bibinfo {author} {\bibfnamefont {M.}~\bibnamefont {Sang}}, \bibinfo {author} {\bibfnamefont {L.}~\bibnamefont {Yuan}},\ and\ \bibinfo {author} {\bibfnamefont {X.}~\bibnamefont {Chen}},\ }\bibfield  {title} {\bibinfo {title} {Multiuser time-energy entanglement swapping based on dense wavelength division multiplexed and sum-frequency generation},\ }\href@noop {} {\bibfield  {journal} {\bibinfo  {journal} {Physical Review Letters}\ }\textbf {\bibinfo {volume} {123}},\ \bibinfo {pages} {250505} (\bibinfo {year} {2019})}\BibitemShut {NoStop}%
	\bibitem [{\citenamefont {Vergyris}\ \emph {et~al.}(2020)\citenamefont {Vergyris}, \citenamefont {Babin}, \citenamefont {Nold}, \citenamefont {Gouzien}, \citenamefont {Herrmann}, \citenamefont {Silberhorn}, \citenamefont {Alibart}, \citenamefont {Tanzilli},\ and\ \citenamefont {Kaiser}}]{vergyris2020}%
		\BibitemOpen
		\bibfield  {author} {\bibinfo {author} {\bibfnamefont {P.}~\bibnamefont {Vergyris}}, \bibinfo {author} {\bibfnamefont {C.}~\bibnamefont {Babin}}, \bibinfo {author} {\bibfnamefont {R.}~\bibnamefont {Nold}}, \bibinfo {author} {\bibfnamefont {E.}~\bibnamefont {Gouzien}}, \bibinfo {author} {\bibfnamefont {H.}~\bibnamefont {Herrmann}}, \bibinfo {author} {\bibfnamefont {C.}~\bibnamefont {Silberhorn}}, \bibinfo {author} {\bibfnamefont {O.}~\bibnamefont {Alibart}}, \bibinfo {author} {\bibfnamefont {S.}~\bibnamefont {Tanzilli}},\ and\ \bibinfo {author} {\bibfnamefont {F.}~\bibnamefont {Kaiser}},\ }\bibfield  {title} {\bibinfo {title} {{Two-photon phase-sensing with single-photon detection}},\ }\bibfield  {journal} {\bibinfo  {journal} {Applied Physics Letters}\ }\textbf {\bibinfo {volume} {117}},\ \href {https://doi.org/10.1063/5.0009527} {10.1063/5.0009527} (\bibinfo {year} {2020}),\ \Eprint {https://arxiv.org/abs/2007.02586} {arXiv:2007.02586} \BibitemShut {NoStop}%
	\bibitem [{\citenamefont {Llewellyn}\ \emph {et~al.}(2020)\citenamefont {Llewellyn}, \citenamefont {Ding}, \citenamefont {Faruque}, \citenamefont {Paesani}, \citenamefont {Bacco}, \citenamefont {Santagati}, \citenamefont {Qian}, \citenamefont {Li}, \citenamefont {Xiao}, \citenamefont {Huber}, \citenamefont {Malik}, \citenamefont {Sinclair}, \citenamefont {Zhou}, \citenamefont {Rottwitt}, \citenamefont {O'Brien}, \citenamefont {Rarity}, \citenamefont {Gong}, \citenamefont {Oxenlowe}, \citenamefont {Wang},\ and\ \citenamefont {Thompson}}]{llewellyn2020}%
		\BibitemOpen
		\bibfield  {author} {\bibinfo {author} {\bibfnamefont {D.}~\bibnamefont {Llewellyn}}, \bibinfo {author} {\bibfnamefont {Y.}~\bibnamefont {Ding}}, \bibinfo {author} {\bibfnamefont {I.~I.}\ \bibnamefont {Faruque}}, \bibinfo {author} {\bibfnamefont {S.}~\bibnamefont {Paesani}}, \bibinfo {author} {\bibfnamefont {D.}~\bibnamefont {Bacco}}, \bibinfo {author} {\bibfnamefont {R.}~\bibnamefont {Santagati}}, \bibinfo {author} {\bibfnamefont {Y.~J.}\ \bibnamefont {Qian}}, \bibinfo {author} {\bibfnamefont {Y.}~\bibnamefont {Li}}, \bibinfo {author} {\bibfnamefont {Y.~F.}\ \bibnamefont {Xiao}}, \bibinfo {author} {\bibfnamefont {M.}~\bibnamefont {Huber}}, \bibinfo {author} {\bibfnamefont {M.}~\bibnamefont {Malik}}, \bibinfo {author} {\bibfnamefont {G.~F.}\ \bibnamefont {Sinclair}}, \bibinfo {author} {\bibfnamefont {X.}~\bibnamefont {Zhou}}, \bibinfo {author} {\bibfnamefont {K.}~\bibnamefont {Rottwitt}}, \bibinfo {author} {\bibfnamefont {J.~L.}\ \bibnamefont {O'Brien}}, \bibinfo {author} {\bibfnamefont {J.~G.}\ \bibnamefont {Rarity}}, \bibinfo {author} {\bibfnamefont {Q.}~\bibnamefont {Gong}}, \bibinfo {author} {\bibfnamefont {L.~K.}\ \bibnamefont {Oxenlowe}}, \bibinfo {author} {\bibfnamefont {J.}~\bibnamefont {Wang}},\ and\ \bibinfo {author} {\bibfnamefont {M.~G.}\ \bibnamefont {Thompson}},\ }\bibfield  {title} {\bibinfo {title} {{Chip-to-chip quantum teleportation and multi-photon entanglement in silicon}},\ }\href {https://doi.org/10.1038/s41567-019-0727-x} {\bibfield  {journal} {\bibinfo  {journal} {Nature Physics}\ }\textbf {\bibinfo {volume} {16}},\ \bibinfo {pages} {148} (\bibinfo {year} {2020})},\ \Eprint {https://arxiv.org/abs/1911.07839} {arXiv:1911.07839} \BibitemShut {NoStop}%
	\bibitem [{\citenamefont {Kneipp}\ \emph {et~al.}(1996)\citenamefont {Kneipp}, \citenamefont {Wang}, \citenamefont {Kneipp}, \citenamefont {Itzkan}, \citenamefont {Dasari},\ and\ \citenamefont {Feld}}]{kneipp1996population}%
		\BibitemOpen
		\bibfield  {author} {\bibinfo {author} {\bibfnamefont {K.}~\bibnamefont {Kneipp}}, \bibinfo {author} {\bibfnamefont {Y.}~\bibnamefont {Wang}}, \bibinfo {author} {\bibfnamefont {H.}~\bibnamefont {Kneipp}}, \bibinfo {author} {\bibfnamefont {I.}~\bibnamefont {Itzkan}}, \bibinfo {author} {\bibfnamefont {R.~R.}\ \bibnamefont {Dasari}},\ and\ \bibinfo {author} {\bibfnamefont {M.~S.}\ \bibnamefont {Feld}},\ }\bibfield  {title} {\bibinfo {title} {Population pumping of excited vibrational states by spontaneous surface-enhanced Raman scattering},\ }\href@noop {} {\bibfield  {journal} {\bibinfo  {journal} {Physical Review Letters}\ }\textbf {\bibinfo {volume} {76}},\ \bibinfo {pages} {2444} (\bibinfo {year} {1996})}\BibitemShut {NoStop}%
	\bibitem [{\citenamefont {Brolo}\ \emph {et~al.}(2004)\citenamefont {Brolo}, \citenamefont {Sanderson},\ and\ \citenamefont {Smith}}]{brolo2004ratio}%
		\BibitemOpen
		\bibfield  {author} {\bibinfo {author} {\bibfnamefont {A.~G.}\ \bibnamefont {Brolo}}, \bibinfo {author} {\bibfnamefont {A.~C.}\ \bibnamefont {Sanderson}},\ and\ \bibinfo {author} {\bibfnamefont {A.~P.}\ \bibnamefont {Smith}},\ }\bibfield  {title} {\bibinfo {title} {Ratio of the surface-enhanced anti-Stokes scattering to the surface-enhanced Stokes-Raman scattering for molecules adsorbed on a silver electrode},\ }\href@noop {} {\bibfield  {journal} {\bibinfo  {journal} {Physical Review B}\ }\textbf {\bibinfo {volume} {69}},\ \bibinfo {pages} {045424} (\bibinfo {year} {2004})}\BibitemShut {NoStop}%
	\bibitem [{\citenamefont {Le~Ru}\ and\ \citenamefont {Etchegoin}(2006)}]{le2006vibrational}%
		\BibitemOpen
		\bibfield  {author} {\bibinfo {author} {\bibfnamefont {E.}~\bibnamefont {Le~Ru}}\ and\ \bibinfo {author} {\bibfnamefont {P.}~\bibnamefont {Etchegoin}},\ }\bibfield  {title} {\bibinfo {title} {Vibrational pumping and heating under sers conditions: fact or myth?},\ }\href@noop {} {\bibfield  {journal} {\bibinfo  {journal} {Faraday Discussions}\ }\textbf {\bibinfo {volume} {132}},\ \bibinfo {pages} {63} (\bibinfo {year} {2006})}\BibitemShut {NoStop}%
	\bibitem [{\citenamefont {Roelli}\ \emph {et~al.}(2016)\citenamefont {Roelli}, \citenamefont {Galland}, \citenamefont {Piro},\ and\ \citenamefont {Kippenberg}}]{roelli2016molecular}%
		\BibitemOpen
		\bibfield  {author} {\bibinfo {author} {\bibfnamefont {P.}~\bibnamefont {Roelli}}, \bibinfo {author} {\bibfnamefont {C.}~\bibnamefont {Galland}}, \bibinfo {author} {\bibfnamefont {N.}~\bibnamefont {Piro}},\ and\ \bibinfo {author} {\bibfnamefont {T.~J.}\ \bibnamefont {Kippenberg}},\ }\bibfield  {title} {\bibinfo {title} {Molecular cavity optomechanics as a theory of plasmon-enhanced Raman scattering},\ }\href@noop {} {\bibfield  {journal} {\bibinfo  {journal} {Nature Nanotechnology}\ }\textbf {\bibinfo {volume} {11}},\ \bibinfo {pages} {164} (\bibinfo {year} {2016})}\BibitemShut {NoStop}%
	\bibitem [{\citenamefont {Schmidt}\ \emph {et~al.}(2016)\citenamefont {Schmidt}, \citenamefont {Esteban}, \citenamefont {Gonz{\'a}lez-Tudela}, \citenamefont {Giedke},\ and\ \citenamefont {Aizpurua}}]{schmidt2016quantum}%
		\BibitemOpen
		\bibfield  {author} {\bibinfo {author} {\bibfnamefont {M.~K.}\ \bibnamefont {Schmidt}}, \bibinfo {author} {\bibfnamefont {R.}~\bibnamefont {Esteban}}, \bibinfo {author} {\bibfnamefont {A.}~\bibnamefont {Gonz{\'a}lez-Tudela}}, \bibinfo {author} {\bibfnamefont {G.}~\bibnamefont {Giedke}},\ and\ \bibinfo {author} {\bibfnamefont {J.}~\bibnamefont {Aizpurua}},\ }\bibfield  {title} {\bibinfo {title} {Quantum mechanical description of Raman scattering from molecules in plasmonic cavities},\ }\href@noop {} {\bibfield  {journal} {\bibinfo  {journal} {ACS nano}\ }\textbf {\bibinfo {volume} {10}},\ \bibinfo {pages} {6291} (\bibinfo {year} {2016})}\BibitemShut {NoStop}%
	\bibitem [{\citenamefont {Zhang}\ \emph {et~al.}(2020)\citenamefont {Zhang}, \citenamefont {Aizpurua},\ and\ \citenamefont {Esteban}}]{zhang2020optomechanical}%
		\BibitemOpen
		\bibfield  {author} {\bibinfo {author} {\bibfnamefont {Y.}~\bibnamefont {Zhang}}, \bibinfo {author} {\bibfnamefont {J.}~\bibnamefont {Aizpurua}},\ and\ \bibinfo {author} {\bibfnamefont {R.}~\bibnamefont {Esteban}},\ }\bibfield  {title} {\bibinfo {title} {Optomechanical collective effects in surface-enhanced Raman scattering from many molecules},\ }\href@noop {} {\bibfield  {journal} {\bibinfo  {journal} {ACS Photonics}\ }\textbf {\bibinfo {volume} {7}},\ \bibinfo {pages} {1676} (\bibinfo {year} {2020})}\BibitemShut {NoStop}%
	\bibitem [{\citenamefont {Kuzmich}\ \emph {et~al.}(2003)\citenamefont {Kuzmich}, \citenamefont {Bowen}, \citenamefont {Boozer}, \citenamefont {Boca}, \citenamefont {Chou}, \citenamefont {Duan},\ and\ \citenamefont {Kimble}}]{Kuzmich2003a}%
		\BibitemOpen
		\bibfield  {author} {\bibinfo {author} {\bibfnamefont {A.}~\bibnamefont {Kuzmich}}, \bibinfo {author} {\bibfnamefont {W.~P.}\ \bibnamefont {Bowen}}, \bibinfo {author} {\bibfnamefont {A.~D.}\ \bibnamefont {Boozer}}, \bibinfo {author} {\bibfnamefont {A.}~\bibnamefont {Boca}}, \bibinfo {author} {\bibfnamefont {C.~W.}\ \bibnamefont {Chou}}, \bibinfo {author} {\bibfnamefont {L.~M.}\ \bibnamefont {Duan}},\ and\ \bibinfo {author} {\bibfnamefont {H.~J.}\ \bibnamefont {Kimble}},\ }\bibfield  {title} {\bibinfo {title} {{Generation of nonclassical photon pairs for scalable quantum communication with atomic ensembles}},\ }\href {https://doi.org/10.1038/nature01714} {\bibfield  {journal} {\bibinfo  {journal} {Nature}\ }\textbf {\bibinfo {volume} {423}},\ \bibinfo {pages} {731} (\bibinfo {year} {2003})},\ \Eprint {https://arxiv.org/abs/0305162} {arXiv:0305162 [quant-ph]} \BibitemShut {NoStop}%
	\bibitem [{\citenamefont {England}\ \emph {et~al.}(2013)\citenamefont {England}, \citenamefont {Bustard}, \citenamefont {Nunn}, \citenamefont {Lausten},\ and\ \citenamefont {Sussman}}]{England2013}%
		\BibitemOpen
		\bibfield  {author} {\bibinfo {author} {\bibfnamefont {D.~G.}\ \bibnamefont {England}}, \bibinfo {author} {\bibfnamefont {P.~J.}\ \bibnamefont {Bustard}}, \bibinfo {author} {\bibfnamefont {J.}~\bibnamefont {Nunn}}, \bibinfo {author} {\bibfnamefont {R.}~\bibnamefont {Lausten}},\ and\ \bibinfo {author} {\bibfnamefont {B.~J.}\ \bibnamefont {Sussman}},\ }\bibfield  {title} {\bibinfo {title} {From photons to phonons and back: A THz optical memory in diamond},\ }\href@noop {} {\bibfield  {journal} {\bibinfo  {journal} {Physical Review Letters}\ }\textbf {\bibinfo {volume} {111}},\ \bibinfo {pages} {243601} (\bibinfo {year} {2013})}\BibitemShut {NoStop}%
	\bibitem [{\citenamefont {Velez}\ \emph {et~al.}(2019)\citenamefont {Velez}, \citenamefont {Seibold}, \citenamefont {Kipfer}, \citenamefont {Anderson}, \citenamefont {Sudhir},\ and\ \citenamefont {Galland}}]{velez2019}%
		\BibitemOpen
		\bibfield  {author} {\bibinfo {author} {\bibfnamefont {S.~T.}\ \bibnamefont {Velez}}, \bibinfo {author} {\bibfnamefont {K.}~\bibnamefont {Seibold}}, \bibinfo {author} {\bibfnamefont {N.}~\bibnamefont {Kipfer}}, \bibinfo {author} {\bibfnamefont {M.~D.}\ \bibnamefont {Anderson}}, \bibinfo {author} {\bibfnamefont {V.}~\bibnamefont {Sudhir}},\ and\ \bibinfo {author} {\bibfnamefont {C.}~\bibnamefont {Galland}},\ }\bibfield  {title} {\bibinfo {title} {Preparation and decay of a single quantum of vibration at ambient conditions},\ }\href@noop {} {\bibfield  {journal} {\bibinfo  {journal} {Physical Review X}\ }\textbf {\bibinfo {volume} {9}},\ \bibinfo {pages} {041007} (\bibinfo {year} {2019})}\BibitemShut {NoStop}%
	\bibitem [{\citenamefont {Vento}\ \emph {et~al.}(2023)\citenamefont {Vento}, \citenamefont {Tarrago~Velez}, \citenamefont {Pogrebna},\ and\ \citenamefont {Galland}}]{vento2023measurement}%
		\BibitemOpen
		\bibfield  {author} {\bibinfo {author} {\bibfnamefont {V.}~\bibnamefont {Vento}}, \bibinfo {author} {\bibfnamefont {S.}~\bibnamefont {Tarrago~Velez}}, \bibinfo {author} {\bibfnamefont {A.}~\bibnamefont {Pogrebna}},\ and\ \bibinfo {author} {\bibfnamefont {C.}~\bibnamefont {Galland}},\ }\bibfield  {title} {\bibinfo {title} {Measurement-induced collective vibrational quantum coherence under spontaneous Raman scattering in a liquid},\ }\href@noop {} {\bibfield  {journal} {\bibinfo  {journal} {Nature Communications}\ }\textbf {\bibinfo {volume} {14}},\ \bibinfo {pages} {2818} (\bibinfo {year} {2023})}\BibitemShut {NoStop}%
	\bibitem [{\citenamefont {Kasperczyk}\ \emph {et~al.}(2016)\citenamefont {Kasperczyk}, \citenamefont {de~Aguiar~J{\'u}nior}, \citenamefont {Rabelo}, \citenamefont {Saraiva}, \citenamefont {Santos}, \citenamefont {Novotny},\ and\ \citenamefont {Jorio}}]{Kasperczyk2016}%
		\BibitemOpen
		\bibfield  {author} {\bibinfo {author} {\bibfnamefont {M.}~\bibnamefont {Kasperczyk}}, \bibinfo {author} {\bibfnamefont {F.~S.}\ \bibnamefont {de~Aguiar~J{\'u}nior}}, \bibinfo {author} {\bibfnamefont {C.}~\bibnamefont {Rabelo}}, \bibinfo {author} {\bibfnamefont {A.}~\bibnamefont {Saraiva}}, \bibinfo {author} {\bibfnamefont {M.~F.}\ \bibnamefont {Santos}}, \bibinfo {author} {\bibfnamefont {L.}~\bibnamefont {Novotny}},\ and\ \bibinfo {author} {\bibfnamefont {A.}~\bibnamefont {Jorio}},\ }\bibfield  {title} {\bibinfo {title} {Temporal quantum correlations in inelastic light scattering from water},\ }\href@noop {} {\bibfield  {journal} {\bibinfo  {journal} {Physical Review Letters}\ }\textbf {\bibinfo {volume} {117}},\ \bibinfo {pages} {243603} (\bibinfo {year} {2016})}\BibitemShut {NoStop}%
	\bibitem [{\citenamefont {Bashkansky}\ \emph {et~al.}(2012)\citenamefont {Bashkansky}, \citenamefont {Fatemi},\ and\ \citenamefont {Vurgaftman}}]{Bashkansky2012}%
		\BibitemOpen
		\bibfield  {author} {\bibinfo {author} {\bibfnamefont {M.}~\bibnamefont {Bashkansky}}, \bibinfo {author} {\bibfnamefont {F.~K.}\ \bibnamefont {Fatemi}},\ and\ \bibinfo {author} {\bibfnamefont {I.}~\bibnamefont {Vurgaftman}},\ }\bibfield  {title} {\bibinfo {title} {{Quantum memory in warm rubidium vapor with buffer gas}},\ }\href {https://doi.org/10.1364/ol.37.000142} {\bibfield  {journal} {\bibinfo  {journal} {Optics Letters}\ }\textbf {\bibinfo {volume} {37}},\ \bibinfo {pages} {142} (\bibinfo {year} {2012})}\BibitemShut {NoStop}%
	\bibitem [{\citenamefont {Reim}\ \emph {et~al.}(2010)\citenamefont {Reim}, \citenamefont {Nunn}, \citenamefont {Lorenz}, \citenamefont {Sussman}, \citenamefont {Lee}, \citenamefont {Langford}, \citenamefont {Jaksch},\ and\ \citenamefont {Walmsley}}]{Reim2010}%
		\BibitemOpen
		\bibfield  {author} {\bibinfo {author} {\bibfnamefont {K.~F.}\ \bibnamefont {Reim}}, \bibinfo {author} {\bibfnamefont {J.}~\bibnamefont {Nunn}}, \bibinfo {author} {\bibfnamefont {V.~O.}\ \bibnamefont {Lorenz}}, \bibinfo {author} {\bibfnamefont {B.~J.}\ \bibnamefont {Sussman}}, \bibinfo {author} {\bibfnamefont {K.~C.}\ \bibnamefont {Lee}}, \bibinfo {author} {\bibfnamefont {N.~K.}\ \bibnamefont {Langford}}, \bibinfo {author} {\bibfnamefont {D.}~\bibnamefont {Jaksch}},\ and\ \bibinfo {author} {\bibfnamefont {I.~A.}\ \bibnamefont {Walmsley}},\ }\bibfield  {title} {\bibinfo {title} {{Towards high-speed optical quantum memories}},\ }\href {https://doi.org/10.1038/nphoton.2010.30} {\bibfield  {journal} {\bibinfo  {journal} {Nature Photonics}\ }\textbf {\bibinfo {volume} {4}},\ \bibinfo {pages} {218} (\bibinfo {year} {2010})},\ \Eprint {https://arxiv.org/abs/0912.2970} {arXiv:0912.2970} \BibitemShut {NoStop}%
	\bibitem [{\citenamefont {Bustard}\ \emph {et~al.}(2015)\citenamefont {Bustard}, \citenamefont {Erskine}, \citenamefont {England}, \citenamefont {Nunn}, \citenamefont {Hockett}, \citenamefont {Lausten}, \citenamefont {Spanner},\ and\ \citenamefont {Sussman}}]{bustard2015}%
		\BibitemOpen
		\bibfield  {author} {\bibinfo {author} {\bibfnamefont {P.~J.}\ \bibnamefont {Bustard}}, \bibinfo {author} {\bibfnamefont {J.}~\bibnamefont {Erskine}}, \bibinfo {author} {\bibfnamefont {D.~G.}\ \bibnamefont {England}}, \bibinfo {author} {\bibfnamefont {J.}~\bibnamefont {Nunn}}, \bibinfo {author} {\bibfnamefont {P.}~\bibnamefont {Hockett}}, \bibinfo {author} {\bibfnamefont {R.}~\bibnamefont {Lausten}}, \bibinfo {author} {\bibfnamefont {M.}~\bibnamefont {Spanner}},\ and\ \bibinfo {author} {\bibfnamefont {B.~J.}\ \bibnamefont {Sussman}},\ }\bibfield  {title} {\bibinfo {title} {{Nonclassical correlations between terahertz-bandwidth photons mediated by rotational quanta in hydrogen molecules}},\ }\href {https://doi.org/10.1364/ol.40.000922} {\bibfield  {journal} {\bibinfo  {journal} {Optics Letters}\ }\textbf {\bibinfo {volume} {40}},\ \bibinfo {pages} {922} (\bibinfo {year} {2015})}\BibitemShut {NoStop}%
	\bibitem [{\citenamefont {Lee}\ \emph {et~al.}(2012)\citenamefont {Lee}, \citenamefont {Sussman}, \citenamefont {Sprague}, \citenamefont {Michelberger}, \citenamefont {Reim}, \citenamefont {Nunn}, \citenamefont {Langford}, \citenamefont {Bustard}, \citenamefont {Jaksch},\ and\ \citenamefont {Walmsley}}]{Lee2012}%
		\BibitemOpen
		\bibfield  {author} {\bibinfo {author} {\bibfnamefont {K.~C.}\ \bibnamefont {Lee}}, \bibinfo {author} {\bibfnamefont {B.~J.}\ \bibnamefont {Sussman}}, \bibinfo {author} {\bibfnamefont {M.~R.}\ \bibnamefont {Sprague}}, \bibinfo {author} {\bibfnamefont {P.}~\bibnamefont {Michelberger}}, \bibinfo {author} {\bibfnamefont {K.~F.}\ \bibnamefont {Reim}}, \bibinfo {author} {\bibfnamefont {J.}~\bibnamefont {Nunn}}, \bibinfo {author} {\bibfnamefont {N.~K.}\ \bibnamefont {Langford}}, \bibinfo {author} {\bibfnamefont {P.~J.}\ \bibnamefont {Bustard}}, \bibinfo {author} {\bibfnamefont {D.}~\bibnamefont {Jaksch}},\ and\ \bibinfo {author} {\bibfnamefont {I.~A.}\ \bibnamefont {Walmsley}},\ }\bibfield  {title} {\bibinfo {title} {{Macroscopic non-classical states and terahertz quantum processing in room-temperature diamond}},\ }\href {https://doi.org/10.1038/nphoton.2011.296} {\bibfield  {journal} {\bibinfo  {journal} {Nature Photonics}\ }\textbf {\bibinfo {volume} {6}},\ \bibinfo {pages} {41} (\bibinfo {year} {2012})}\BibitemShut {NoStop}%
	\bibitem [{\citenamefont {Kasperczyk}\ \emph {et~al.}(2015)\citenamefont {Kasperczyk}, \citenamefont {Jorio}, \citenamefont {Neu}, \citenamefont {Maletinsky},\ and\ \citenamefont {Novotny}}]{kasperczyk2015stokes}%
		\BibitemOpen
		\bibfield  {author} {\bibinfo {author} {\bibfnamefont {M.}~\bibnamefont {Kasperczyk}}, \bibinfo {author} {\bibfnamefont {A.}~\bibnamefont {Jorio}}, \bibinfo {author} {\bibfnamefont {E.}~\bibnamefont {Neu}}, \bibinfo {author} {\bibfnamefont {P.}~\bibnamefont {Maletinsky}},\ and\ \bibinfo {author} {\bibfnamefont {L.}~\bibnamefont {Novotny}},\ }\bibfield  {title} {\bibinfo {title} {{Stokes–anti-Stokes correlations in diamond}},\ }\href {https://doi.org/10.1364/ol.40.002393} {\bibfield  {journal} {\bibinfo  {journal} {Optics Letters}\ }\textbf {\bibinfo {volume} {40}},\ \bibinfo {pages} {2393} (\bibinfo {year} {2015})}\BibitemShut {NoStop}%
	\bibitem [{\citenamefont {Anderson}\ \emph {et~al.}(2018)\citenamefont {Anderson}, \citenamefont {Velez}, \citenamefont {Seibold}, \citenamefont {Flayac}, \citenamefont {Savona}, \citenamefont {Sangouard},\ and\ \citenamefont {Galland}}]{anderson2018}%
		\BibitemOpen
		\bibfield  {author} {\bibinfo {author} {\bibfnamefont {M.~D.}\ \bibnamefont {Anderson}}, \bibinfo {author} {\bibfnamefont {S.~T.}\ \bibnamefont {Velez}}, \bibinfo {author} {\bibfnamefont {K.}~\bibnamefont {Seibold}}, \bibinfo {author} {\bibfnamefont {H.}~\bibnamefont {Flayac}}, \bibinfo {author} {\bibfnamefont {V.}~\bibnamefont {Savona}}, \bibinfo {author} {\bibfnamefont {N.}~\bibnamefont {Sangouard}},\ and\ \bibinfo {author} {\bibfnamefont {C.}~\bibnamefont {Galland}},\ }\bibfield  {title} {\bibinfo {title} {Two-color pump-probe measurement of photonic quantum correlations mediated by a single phonon},\ }\href@noop {} {\bibfield  {journal} {\bibinfo  {journal} {Physical Review Letters}\ }\textbf {\bibinfo {volume} {120}},\ \bibinfo {pages} {233601} (\bibinfo {year} {2018})}\BibitemShut {NoStop}%
	\bibitem [{\citenamefont {J{\'u}nior}\ \emph {et~al.}(2020)\citenamefont {J{\'u}nior}, \citenamefont {Santos}, \citenamefont {Monken},\ and\ \citenamefont {Jorio}}]{Junior2020}%
		\BibitemOpen
		\bibfield  {author} {\bibinfo {author} {\bibfnamefont {F.~S.}\ \bibnamefont {de~Aguiar~J{\'u}nior}}, \bibinfo {author} {\bibfnamefont {M.~F.}\ \bibnamefont {Santos}}, \bibinfo {author} {\bibfnamefont {C.~H.}\ \bibnamefont {Monken}},\ and\ \bibinfo {author} {\bibfnamefont {A.}~\bibnamefont {Jorio}},\ }\bibfield  {title} {\bibinfo {title} {Lifetime and polarization for real and virtual correlated Stokes-anti-Stokes Raman scattering in diamond},\ }\href@noop {} {\bibfield  {journal} {\bibinfo  {journal} {Physical Review Research}\ }\textbf {\bibinfo {volume} {2}},\ \bibinfo {pages} {013084} (\bibinfo {year} {2020})}\BibitemShut {NoStop}%
	\bibitem [{\citenamefont {Lee}\ \emph {et~al.}(2011)\citenamefont {Lee}, \citenamefont {Sprague}, \citenamefont {Sussman}, \citenamefont {Nunn}, \citenamefont {Langford}, \citenamefont {Jin}, \citenamefont {Champion}, \citenamefont {Michelberger}, \citenamefont {Reim}, \citenamefont {England}, \citenamefont {Jaksch},\ and\ \citenamefont {Walmsley}}]{Lee2011a}%
		\BibitemOpen
		\bibfield  {author} {\bibinfo {author} {\bibfnamefont {K.~C.}\ \bibnamefont {Lee}}, \bibinfo {author} {\bibfnamefont {M.~R.}\ \bibnamefont {Sprague}}, \bibinfo {author} {\bibfnamefont {B.~J.}\ \bibnamefont {Sussman}}, \bibinfo {author} {\bibfnamefont {J.}~\bibnamefont {Nunn}}, \bibinfo {author} {\bibfnamefont {N.~K.}\ \bibnamefont {Langford}}, \bibinfo {author} {\bibfnamefont {X.~M.}\ \bibnamefont {Jin}}, \bibinfo {author} {\bibfnamefont {T.}~\bibnamefont {Champion}}, \bibinfo {author} {\bibfnamefont {P.}~\bibnamefont {Michelberger}}, \bibinfo {author} {\bibfnamefont {K.~F.}\ \bibnamefont {Reim}}, \bibinfo {author} {\bibfnamefont {D.}~\bibnamefont {England}}, \bibinfo {author} {\bibfnamefont {D.}~\bibnamefont {Jaksch}},\ and\ \bibinfo {author} {\bibfnamefont {I.~A.}\ \bibnamefont {Walmsley}},\ }\bibfield  {title} {\bibinfo {title} {{Entangling macroscopic diamonds at room temperature}},\ }\href {https://doi.org/10.1126/science.1211914} {\bibfield  {journal} {\bibinfo  {journal} {Science}\ }\textbf {\bibinfo {volume} {334}},\ \bibinfo {pages} {1253} (\bibinfo {year} {2011})}\BibitemShut {NoStop}%
	\bibitem [{\citenamefont {Velez}\ \emph {et~al.}(2020)\citenamefont {Velez}, \citenamefont {Sudhir}, \citenamefont {Sangouard},\ and\ \citenamefont {Galland}}]{velez2020}%
		\BibitemOpen
		\bibfield  {author} {\bibinfo {author} {\bibfnamefont {S.~T.}\ \bibnamefont {Velez}}, \bibinfo {author} {\bibfnamefont {V.}~\bibnamefont {Sudhir}}, \bibinfo {author} {\bibfnamefont {N.}~\bibnamefont {Sangouard}},\ and\ \bibinfo {author} {\bibfnamefont {C.}~\bibnamefont {Galland}},\ }\bibfield  {title} {\bibinfo {title} {{Bell correlations between light and vibration at ambient conditions}},\ }\bibfield  {journal} {\bibinfo  {journal} {Science Advances}\ }\textbf {\bibinfo {volume} {6}},\ \href {https://doi.org/10.1126/SCIADV.ABB0260} {10.1126/SCIADV.ABB0260} (\bibinfo {year} {2020}),\ \Eprint {https://arxiv.org/abs/1912.04502} {arXiv:1912.04502} \BibitemShut {NoStop}%
	\bibitem [{\citenamefont {Dolde}\ \emph {et~al.}(2013)\citenamefont {Dolde}, \citenamefont {Jakobi}, \citenamefont {Naydenov}, \citenamefont {Zhao}, \citenamefont {Pezzagna}, \citenamefont {Trautmann}, \citenamefont {Meijer}, \citenamefont {Neumann}, \citenamefont {Jelezko},\ and\ \citenamefont {Wrachtrup}}]{dolde2013}%
		\BibitemOpen
		\bibfield  {author} {\bibinfo {author} {\bibfnamefont {F.}~\bibnamefont {Dolde}}, \bibinfo {author} {\bibfnamefont {I.}~\bibnamefont {Jakobi}}, \bibinfo {author} {\bibfnamefont {B.}~\bibnamefont {Naydenov}}, \bibinfo {author} {\bibfnamefont {N.}~\bibnamefont {Zhao}}, \bibinfo {author} {\bibfnamefont {S.}~\bibnamefont {Pezzagna}}, \bibinfo {author} {\bibfnamefont {C.}~\bibnamefont {Trautmann}}, \bibinfo {author} {\bibfnamefont {J.}~\bibnamefont {Meijer}}, \bibinfo {author} {\bibfnamefont {P.}~\bibnamefont {Neumann}}, \bibinfo {author} {\bibfnamefont {F.}~\bibnamefont {Jelezko}},\ and\ \bibinfo {author} {\bibfnamefont {J.}~\bibnamefont {Wrachtrup}},\ }\bibfield  {title} {\bibinfo {title} {{Room-temperature entanglement between single defect spins in diamond}},\ }\href {https://doi.org/10.1038/nphys2545} {\bibfield  {journal} {\bibinfo  {journal} {Nature Physics}\ }\textbf {\bibinfo {volume} {9}},\ \bibinfo {pages} {139} (\bibinfo {year} {2013})}\BibitemShut {NoStop}%
	\end{thebibliography}

%

\end{document}